\documentclass[aps,prl,superscriptaddress,twocolumn,floats,tightenlines,balancelastpage,floatfix]{revtex4}
\input epsf
\usepackage{amsmath,amssymb}
\usepackage{epsfig}
\usepackage{graphicx}

\def\mathrelfun#1#2{\lower3.6pt\vbox{\baselineskip0pt\lineskip.9pt
  \ialign{$\mathsurround=0pt#1\hfil##\hfil$\crcr#2\crcr\sim\crcr}}}
\def\simlt{\mathrel{\mathpalette\mathrelfun <}}
\def\simgt{\mathrel{\mathpalette\mathrelfun >}}

\def\ln {{\rm ln}}
\def\hatbfn{{\hat{\bf n}}}

\def\bfx{{\bf x}}
\def\RV{{R_{\rm V}}}
\def\zV{{z_{\rm edge}}}
\def\zrei{{z_{\rm rei}}}
\def\zi{{z_{\rm i}}}
\def\ze{{z_{\rm e}}}
\def\DA{{D_{\rm A}}}
\def\Heo{{H_{{\rm e}0}}}

\begin{document}

\title{A Test of the Copernican Principle}

\author{R.~R. Caldwell}
\affiliation{Department of Physics \& Astronomy, 6127 Wilder Lab, Dartmouth College, 
Hanover, NH 03755}

\author{A. Stebbins}
\affiliation{Theoretical Astrophysics Group, Fermi National Accelerator
Laboratory, P.O. Box 500, Batavia, IL 60510}
  
\date{\today}

\begin{abstract}

The blackbody nature of the cosmic microwave background (CMB) radiation spectrum is
used in a modern test of the Copernican Principle. The reionized universe serves as
a mirror to reflect CMB photons, thereby permitting a view of ourselves and the
local gravitational potential. By comparing with measurements of the CMB spectrum, a
limit is placed on the possibility that we occupy a privileged location, residing at
the center of a large void. The Hubble diagram inferred from lines-of-sight
originating at the center of the void may be misinterpreted to indicate cosmic
acceleration. Current limits on spectral distortions are shown to exclude the
largest voids which mimic cosmic acceleration. More sensitive measurements of the
CMB spectrum could prove the existence of such a void or confirm the validity of the
Copernican Principle.

\end{abstract}

\maketitle


{\it Introduction:} The observed accelerating expansion of the universe
\cite{Riess:1998cb,Perlmutter:1998np} poses deep questions for cosmology. Is the
universe filled by some new, exotic dark energy, or is one of the basic tenets of
the standard model of cosmology invalid? One such tenet is the Cosmological
Principle, the assumption of approximate homogeneity and isotropy of matter and
radiation throughout the universe. The Cosmological Principle is known to be partly
satisfied. The universe is observed to be very nearly isotropic on our celestial
sphere, on the basis of the near-isotropy of the CMB temperature pattern
\cite{CMBisotropy}. The universe is observed to be approximately
homogeneous across the distances probed by large-scale structures
\cite{Hogg:2004vw}. Yet, radial homogeneity on cosmic scales $\simgt1\,$Gpc remains
to be proven. If the assumption of radial homogeneity is relaxed, and if we observe
from a preferred vantage point, then it may be possible to explain the apparent
cosmic acceleration in terms of a peculiar distribution of matter centered upon our
location 
\cite{RefGroupA}. In fact, models of the universe
consisting of a spherically-symmetric distribution of matter, mathematically
described by a Lemaitre-Tolman-Bondi spacetime \cite{Bondi:1947av}, have been shown
to produce a Hubble diagram which is consistent with observations.  These models
require no cosmological constant or other form of dark energy,  and locally resemble
a matter-dominated low-density universe or void. The observed near-isotropy
constrains us to occupy a very special location, at or near the center of the void,
in violation of the Copernican Principle.  Although the Copernican Principle may be
widely accepted by {\it fiat}, it is imperative that such a foundational principle
be proven.  

We propose a test of the Copernican Principle, to verify radial homogeneity and
thereby constrain non-accelerating void cosmological models. The test relies on a
previously under-appreciated effect: the mixture of anisotropic CMB radiation
through scattering leads to distortions of the blackbody spectrum
\cite{Stebbins2007}. The CMB is initially thermal (blackbody), but small
inhomogeneities cause variations in the temperature at different locations and along
different lines-of-sight that preserve the blackbody spectrum.  However, scattering
of this anisotropic radiation into our line-of-sight by ionized gas produces
observable spectral distortions. This allows us to indirectly detect large
anisotropies in other parts of the universe. 

\begin{figure}[h]
\centerline{\epsfig{file=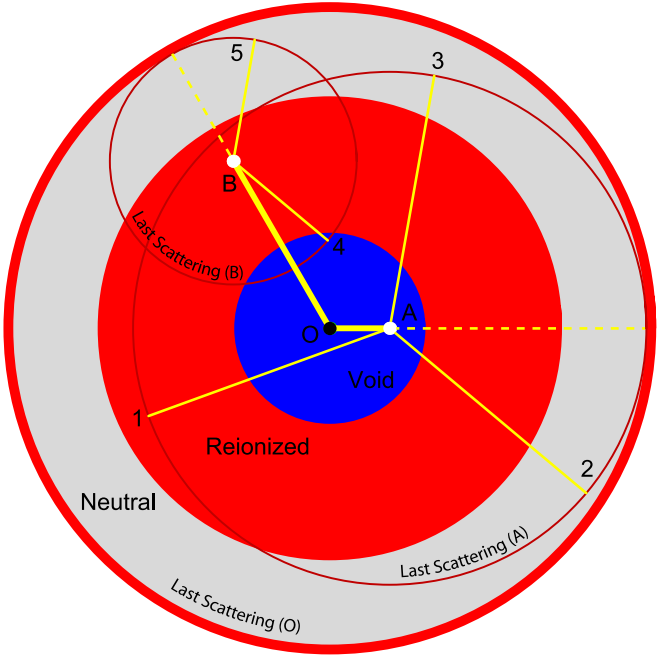,width=20pc,angle=0}}
\caption{Illustrated is a cross-section through a model universe with the observer
(O) at the center of a void, in violation of the Copernican Principle. CMB photons
traveling in any direction may Thomson scatter off reionized gas toward the observer. 
The final spectrum of the observed light will be a mixture of blackbody spectra with
different (anisotropic) tempertures, producing a distorted blackbody. The
yellow lines represent: incoming beams of unscattered, primary CMB photons (dashed);
incoming beams of scattered photons (thin), and the observed beams (thick) for
representative scattering centers with last scattering surfaces represented by
the dark circles. A is in the {\it Doppler zone}: Beams 1-3 experience the same SW
temperature shift, introducing no anisotropy. However,
gradients in the void gravitational potential cause the gas to move with respect to
the CMB frame, so A sees a differential Doppler anisotropy, resulting in spectral
distortions. B is in the {\it reflection zone}: B is at rest with respect to 
the CMB frame and sees no Doppler anisotropy. However, some of the incoming photons, 
{\it e.g.} beam 4, originate inside the void so there will be an anisotropic SW
temperature shift, leading to spectral distortions. 
}
\label{fig:bubble}
\vspace{-0.3cm}
\end{figure}

Here we are interested in anisotropies caused by a large, local void.  Such a 
structure causes ionized gas to move outward, in motion relative to the CMB frame 
which leads to a Doppler anisotropy in the gas frame. The gravitational potential of
such a structure also leads to a Sachs-Wolfe (SW) effect for photons which originate
inside of the void and scatter back toward us.  The geometry of these  effects is
illustrated in fig.~\ref{fig:bubble}. A large void, or any other non-Copernican
structure, will lead to large anisotropies in other places which  will be reflected
back at us in the form of spectral distortions. Hence, deviations from a blackbody
spectrum can indicate a violation of the Copernican Principle. In essence, we use
the reionized universe as a mirror to look at ourselves in CMB light.  If we see
ourselves in the the mirror it is because ours is a privileged location. If we see
nothing in the mirror,  then the Copernican Principle is upheld. 

{\it Spectral Distortions:}
The distortion of the CMB blackbody spectrum due to scattering by anisotropic CMB
radiation is \cite{Stebbins2007} 
$u[\hatbfn]={3\over16\pi}\int_0^\infty
dz\,{d\tau\over dz}\, \int
d^2\hatbfn'\,\left(1+(\hatbfn\cdot\hatbfn')^2\right)$
$\times$ 
$\left( {\Delta T\over T}[\hatbfn ,\hatbfn,z] -{\Delta T\over
T}[\hatbfn',\hatbfn,z]\right)^2,
$
where $\Delta T/T[\hatbfn',\hatbfn,z]$ is the CMB temperature anisotropy in the
direction $\hatbfn'$, as observed at redshift $z$ in the direction $\hatbfn$ from
the central observer, and $\tau$ is the optical depth. For cosmic voids extending
out to redshifts $z\simlt1$, reflections back at us may occur up to $z\simlt3$
(see fig.~\ref{fig:bubble}).  The optical depth to Thomson scattering is small,
so that it is appropriate to consider single scattering. Since the mean CMB
temperature is not known {\it a priori}, but rather is fit to the observations,  $u$
is observationally degenerate with the Compton $y$-distortion parameter according to
the relation $u=2y$. (Compare Refs.~\cite{Stebbins2007,Chluba:2003} for details.)
Thus observational constraints on $2y$ can can be treated as constraints on $u$.

We consider a low-amplitude void embedded in a flat, Einstein-deSitter (EdS or 
$\Omega=1$) matter-dominated universe. The gravitational potential due to
the void, $\Phi[\bfx]$, is a function of comoving position, $\bfx$, with Earth near
$\bfx=0$. The temperature anisotropy can be divided into a Sachs-Wolfe and Doppler
term 
${\Delta T\over T}[\hatbfn',\hatbfn,z]=
{\Delta T\over T}|_{\rm SW}+{\Delta T\over T}|_{\rm Doppler}$
where
${\Delta T\over T}|_{\rm SW}=
\frac{1}{3 c^2}(\Phi[\bfx_{\rm rec}]-\Phi[\bfx_{\rm scatter}])$
and 
${\Delta T\over T}|_{\rm Doppler}=\frac{2}{3} 
{\hatbfn'\cdot{\bf \nabla}_\bfx\Phi[\bfx_{\rm scatter}]}/{c\,H_0\sqrt{1+z}}$,
where
$\bfx_{\rm scatter}= D_{\rm A}^{\rm co}[z]\,\hatbfn$, 
$\bfx_{\rm rec}= \bfx_{\rm scatter}
+(D_{\rm A}^{\rm co}[z_{\rm rec}]-D_{\rm A}^{\rm co}[z])\hatbfn'$,
$D_{\rm A}^{\rm co}[z]=2{c\over\Heo}\,\left(1-{1\over\sqrt{1+z}}\right)$.
Here $D_{\rm A}^{\rm co}$  is the comoving angular diameter distance, and the
redshift of recombination, $z_{\rm rec}$, will be approximated by $\infty$ for
simplicity.  The Hubble constant at the present time in the background cosmology,
outside the void, is $\Heo$, whereas $H_0$ is the larger, present-day Hubble
constant at the center of the void.

We neglect the integrated Sachs-Wolfe (ISW) effect, meaning that a CMB photon does
not contribute to the $u$-distortion simply because it passes across the void. This
approximation is justified for a low-amplitude void in the EdS background where the
ISW is a second-order effect. As $\Omega$ deviates from unity and/or the void
amplitude becomes non-linear we expect a larger ISW contribution to the anisotropy
and thus to the spectral distortion, but we do not expect that the ISW will ever be
the dominant contributor to $u$ for the small voids needed to mimic an accelerating
universe.


The run of optical depth with redshift is taken from the unperturbed, background
cosmology. We assume a rapid reionization at $z=\zrei$ such that
${d\tau\over dz}=\tau_{\rm e0}' \, \sqrt{1+z}\,\Theta[\zrei-z]$
$\tau_{\rm e0}' = {3\Heo\Omega_{\rm b0}\sigma_{\rm T} c\over8\pi G m_{\rm H}}\,
\left(1-{1\over2}Y_{\rm He}\right)$,
where $\Theta[x]$ is the Lorentz-Heaviside step function, $\sigma_{\rm T},\, m_{\rm
H},\, \Omega_{\rm b0}$, and $Y_{\rm He}$ are the Thomson cross-section, the hydrogen
mass, the current baryonic mass density in units of the critical density, and the
helium mass fraction, respectively.  We use $\Omega_{\rm b0}h^2=0.022$
($h\equiv\Heo/100$km/s/Mpc), $Y_{\rm He}=0.24$.   For $H_0$ we use the 
locally-measured expansion rate: $73$~km/s/Mpc ({\it e.g.}
Refs.~\cite{Freedman:2000cf,Riess:2005zi}). Where needed we use the WMAP3 
\cite{SpergelBean2006} value, $\tau_{\rm obs}=0.9$, for the optical depth to the
surface of last-scattering which in our model gives $\zrei=11$. These numbers
specify the cosmic evolution of the density of scatterers.  


We assume spherical symmetry for the local void. Consequently, the gravitational
potential is $\Phi[\bfx]=\Phi[R=|\bfx|]$, where $R$ is the comoving radial distance
from Earth. The temperature anisotropy $\Delta T/T$ depends on the
directions $\hatbfn$ and $\hatbfn'$ only through the combination
$\hatbfn\cdot\hatbfn'$, which leaves $u$ $\hatbfn$-independent.  Thus the final
result is a single number, the $u$-distortion at Earth, which can be translated into
a limit on any local spherical inhomogeneity.

{\it Void Model:}
We cannot compute $u$ for every possible void profile, so we focus our attention on
a particularly simple, two parameter class of voids, sometimes known as a
{\it Hubble bubble}:
$\Phi[R]=\Phi_0\,\left(1-{R^2\over R_{\rm V}^2}\right)\,{\Theta}[\RV-R]$.
The parameters $\Phi_0,\,\RV$ give the void amplitude and comoving radius. The
reason it is called a Hubble bubble is that the Hubble parameter
is uniform inside and outside the void, but the values differ. Nonlinear growth
leads to the appearance of a shell of mass overdensity which compensates the
underdensity in the void at the boundary of the outer and inner region.  This
compensating shell has a complicated density and velocity structure, which is safely
ignored in linear theory. Away from the compensating shell this model resembles an
open ($\Omega_0<1$) FRW cosmology embedded inside a flat EdS cosmology.  Any smooth
spherical void which is asymptotically EdS at large $R$ and has finite density in
the center can be thought of in this way; what differs is the radial profile of the
transition between the two FRW spacetimes.  The Hubble bubble is the limit of a
sharp transition between the interior and exterior regions.

\begin{figure}
\centerline{\epsfig{file=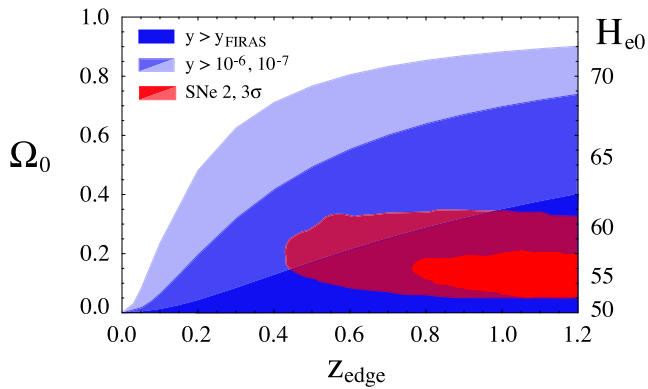,width=20pc,angle=0}}
\caption{The test of the Copernican Principle, in terms of constraints on the
size and depth of a local, spherically-symmetric void, is shown. The blue shaded
regions show the range of parameters excluded by the $u$-distortion test, whereas 
the red regions show the range of parameters compatible with the current SNe Hubble
diagram data.}
\label{HBfig}
\vspace{-0.3cm}
\end{figure}

The Hubble bubble amplitude can be expressed in terms of the present-day density
parameter, $\Omega_0$, inside the void as
$\Phi_0={3\over20}\,(\Heo\RV)^2
\left[\frac{(1-\Omega_0)^{3/2}}{\Omega_0}\frac{H_0}{\Heo}\right]^{2/3}$.
Next, the void radius can be expressed as a function of the redshift at the edge of
the void, $\zV$.  This relationship is complicated by finite peculiar velocities and
non-linear clustering of the compensating shell, but to first order is simply
$\RV=2(c/ H_0)(1-{1/\sqrt{1+\zV}})$. Finally, the exterior Hubble parameter, $\Heo$,
differs from the interior value, $H_0$.  At the same ``time since bang" they are
related as
$\frac{H_0}{\Heo}={3\over2}{\sqrt{1-\Omega_0}-\Omega_0\,
{\rm sinh}^{-1}\left[\sqrt{1-\Omega_0\over\Omega_0}\right]
\over(1-\Omega_0)^{3/2}}$.
Note that a small jump in the Hubble parameter corresponds to a large jump 
in the density parameter (see fig.~\ref{HBfig}).


The gas at different redshift must satisfy several criteria in order to contribute
to the $u$-distortion. A patch of gas at redshift $z$ must be ionized, on our past
light cone, and see an anisotropic $\Delta T$ from the void.  We refer to the region
$z>\zrei$ as the {\it neutral zone} because the gas is not ionized, producing no
contribution to $u$. Even if the gas is ionized, if $z> z_{\rm max} \equiv 3+4\,\zV$
then the gas is not in causal contact with the void so $\Delta T=0$. We refer to the
range  $(2\sqrt{z_{\rm max}+1}-1)/(\sqrt{z_{\rm max}+1}-1)^2 \le z \le z_{\rm max}$
as the {\it reflection zone};  the last scattering surface of gas in this range intersects
the interior of the void so that, depending on the scattering angle, some CMB photons
will reflect back towards us with anisotropy $\Delta T_{\rm SW}$. Gas which is on
our past light cone and within the void will see $\Delta T_{\rm Doppler}$, which we
call the {\it Doppler zone}.

\begin{figure}
\centerline{\epsfig{file=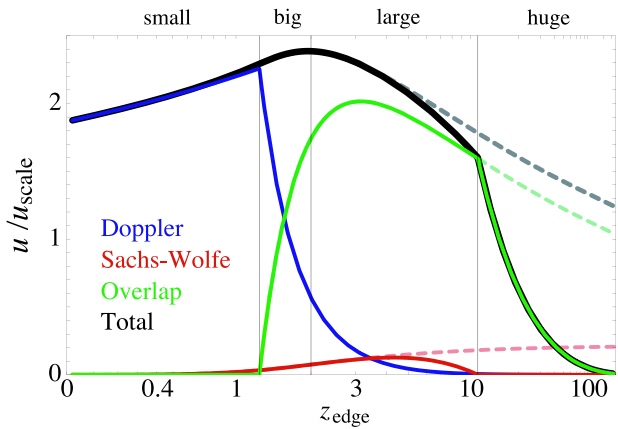,width=19pc,angle=0}}
\caption{The dependence of the spectral distortion, $u$, on the size of a Hubble
bubble parameterized by $\zV$, is shown in units of $u_{\rm scale}\equiv\tau_{\rm
e0}'\,(\Phi_0/(3c^2))^2(1+\zV)/(\sqrt{1+\zV}-1)$.  The thick curves show the
various contributions   to $u$. The  dashed curves correspond to the case in
which $\zrei\to\infty$. For small voids the Doppler  contribution
dominates, and the value of $\zrei$ is unimportant.}
\label{uofzv}
\vspace{-0.3cm}
\end{figure}

Five classes of void sizes are identified depending on how the different zones
overlap (assuming $\zrei > 8$): {\it small} ($\zV\le{5\over4}$) whereby the neutral
zone, reflection zone, and Doppler zone are all disjoint; {\it big}
(${5\over4}<\zV\le{1\over4}(\zrei-3)$) whereby the Doppler zone and reflection zones
overlap, but neither overlap the neutral zone; {\it large}
(${1\over4}(\zrei-3)<\zV\le\zrei$) in which the Doppler and reflection zones
overlap, as do the reflection and neutral zones, but the neutral zone does not
overlap the Doppler zone; {\it huge} ($\zV>\zrei$) in which the neutral, reflection,
and Doppler zones all overlap; and {\it super-horizon} ($\RV>2c/H_0$) for which the
void encompasses the entire observable universe. This classification is not
restricted to the Hubble bubble void profile, but applies to any void profile with a
sharp edge at $z=\zV$.  As we shall see it is only the small voids that can explain
the current SNe data.


In the linear perturbation approximation for this void model the spectral distortion
$u$ is proportional to $({1\over3c^2}\Phi_0)^2\tau_{\rm e0}$ and may be decomposed
as
$u=\tau_{\rm e0}\,\left({\Phi_0\over3c^2}\right)^2\,
\left(U_{\rm D}+U_{\rm S}+U_{\rm DS}\right)$
where the three terms are, respectively, the contribution from gas where the
temperature anisotropies are Doppler only (subscript D), Sachs-Wolfe only 
(subscript S), and a combination of the two (subscript DS). All of these can be
expressed analytically. For small and big voids $u$ does not depend on $\zrei$ but
only on the  dimensionless size parameter $r\equiv{1\over2}{H_0\over
c}\RV=1-{1\over\sqrt{1+z_V}}$. The general expression for $u$ is long and we do not
give it here. For small voids, which are the most relevant, we find
$U_{\rm D}^{\rm small}={28\over5}{1\over r^3}
\left(1+{1\over1-r}+{2\over r}\ln[1-r]\right)$
and $U_{\rm S}^{\rm small}\ll U_{\rm D}^{\rm small}$ and $U_{\rm DS}^{\rm small}=0$.

 
The angular-diameter distance $\DA$ is a solution of the Dyer-Roeder 
\cite{DyerRoeder} equation, $\frac{d}{dz}\left((1+z)^2 H \frac{d}{dz}\DA\right) +
\frac{3}{2}\Omega H \DA=0$.  In the interior open and exterior flat cosmologies the
respective solutions are
$\DA[z<\zi]={2c\over H_0} 
({2-\Omega_0(1-z)-(2-\Omega_0)\sqrt{1+z\Omega_0})/(1+z)^2 \Omega_0^2} $ and
$\DA[z>\ze]={2c\over H_0}(\frac{C_1}{(1+z)} + \frac{C_2}{(1+z)^{3/2}})$,
where the coefficients $C_1,\,C_2$ are set by the continuity $\DA^{\rm
int}[\zi]=\DA^{\rm ext}[\ze]$, and the jump in $d\DA/dz$ as determined by integrating
the Dyer-Roeder equation across the delta-function density spike at the void edge.
The  radial velocity drop, $\Delta v$, at the void edge means a double-valued
$\DA[z]$ for $z\in[\ze,\zi]$ and ${1+\zi\over1+\ze}=\sqrt{c+\Delta v\over c-\Delta
v}$.  This drop also gives the Doppler anisotropy at the edge. To get $\zi$  and
$\ze$ we use the approximations ${\Delta v\over c}={\Delta T\over T}_{\rm
Doppler}[\zV,\hatbfn,\hatbfn]$ and $\zV={1\over2}(\zi+\ze)$.  The luminosity distance
versus redshift, a.k.a. the Hubble diagram, is $(1+z)^2\,D_{\rm A}[z]$. 

{\it Constraints:}
The $u$-distortion is evaluated according to the procedure described above. We are
primarily interested in small and big voids which extend out to $z\sim 1$. Hence our
constraints are independent of $\zrei$.  The other cosmological parameters only
enter into the overall normalization of $u$ through $\tau_{\rm e0}' $.  What remains
are the void, size and amplitude:  ($\zV,\,\Omega_0$).

The  best current bound on $u$ is due to FIRAS
\cite{Mather:1994,BoggessMatherEtAl1992,FixsenCheng1996} which constrains
$y<15\times10^{-6}$ or $u<3\times10^{-5}$ at $95\%$ C.L..  The corresponding 
constraint on Hubble bubble parameters are shown in fig.~\ref{HBfig}.   Also shown
are constraints for projected bounds $y<10^{-6},\, 10^{-7}$.  The limits are
expected to improve  \cite{Fixsen:2002,Kogut:2006kf}, but a $y$-distortion from the
IGM would likely mask the signal discussed here if $u\simlt10^{-6}$
\cite{ZhangPenTrac}.

The results rule out large voids with large density contrasts -- the most egregious
violations of the Copernican Principle.  The larger the void, the smaller the
density contrast must be in order to pass the test.  Although not shown, the
constraints become weaker for huge (nearly super-horizon sized) voids.   Since
observationally $\Omega_0 \lesssim 0.3$, only small bubbles with $\zV <0.9$ are
allowed. Improving the constraint to $y<10^{-6}$ would lower this bound to
$\zV\lesssim0.3$ or a radius of $1$~Gpc. These constraints are consistent with the
very small Hubble bubble proposed in Ref.~\cite{Jha:2006fm}, with $H_0-\Heo \sim0.1
H_0$ and  $\zV\simgt0.025$.

The observed SNe data can be compared with our model Hubble diagram to further
constrain void parameters. Using the SNe data \cite{Riess:2006fw,WoodVasey:2007jb}
compiled in Ref.~\cite{Davis:2007na}, we computed the likelihood of $\Omega_0$ and
$\zV$. The best-fit parameter combinations give $\chi^2=207$ for the $192$ SNe
magnitudes (within $3\sigma$ of the best-fit $\Lambda$CDM model based on a
$\Delta\chi^2$ test, for a family of models with a sufficient number of parameters to
encompass both $\Lambda$ and the void). Voids which explain the observed Hubble
diagram have low density and large size, $\zV\sim 1$ (radii $\sim 2.5$~Gpc). However,
as shown in fig.~\ref{HBfig}, combining the SNe data with current limits on $u$
($\chi^2=225,\,250$ for $191$ degrees of
freedom), we find that nearly all such voids are ruled out. These specific
constraints only apply to  the Hubble bubble class of models, which also suffers from
other flaws not mentioned here. This test will be applied to more general and more
realistic void profiles.

An improvement in the bound on $u$ by an order of magnitude may confirm or refute a
wider variety of such voids as an explanation of the dark energy phenomena.  Yet, the
$u$-distortion test presented here is more general than the question of dark energy.
Future pursuit of this test will help improve our view of the universe on the largest
scales.

\begin{acknowledgments} 

We thank the Galileo Galilei Institute for Theoretical Physics for the hospitality
and the INFN for partial support during the completion of this work. R.C. was
supported in part by NSF AST-0349213 at Dartmouth and A.S. by the DoE at
Fermilab. 
\vspace{-0.5cm}
\end{acknowledgments}


\vfill

\end{document}